\newcommand{\fp}{Fabry-P\'erot }  
\documentclass[epj]{svjour}
\usepackage{graphics}
\usepackage{color}
\usepackage{cite}
\begin{document}
\title{The OVAL experiment: A new experiment to measure vacuum magnetic birefringence using high repetition pulsed magnets
}
\author{Xing Fan\inst{1}
\thanks{\emph{Corresponding author:} xfan@icepp.s.u-tokyo.ac.jp}%
 \and Shusei Kamioka\inst{1}
 \and Toshiaki Inada\inst{2}
 \and Takayuki Yamazaki\inst{2}
 \and Toshio Namba\inst{2}
 \and Shoji Asai\inst{1}
 \and Junko Omachi\inst{3}
 \and Kosuke Yoshioka\inst{4}
 \and Makoto Kuwata-Gonokami\inst{1}
 \and Akira Matsuo\inst{5}
 \and Koushi Kawaguchi\inst{5}
 \and Koichi Kindo\inst{5}
 \and Hiroyuki Nojiri\inst{6}
 }                     
%
%
\institute{Department of Physics, Graduate School of Science, The University of Tokyo, 7-3-1 Hongo, Bunkyo-ku, Tokyo 113-0033 Japan
 \and International Center for Elementary Particle Physics, The University of Tokyo, 7-3-1 Hongo, Bunkyo-ku, Tokyo 113-0033, Japan
 \and Institute for Photon Science and Technology, The University of Tokyo, 7-3-1 Hongo, Bunkyo-ku, Tokyo 113-0033, Japan
 \and Photon Science Center (PSC), The University of Tokyo, 2-11-16 Yayoi, Bunkyo-ku, Tokyo 113-8658, Japan
 \and The Institute for Solid State Physics, The University of Tokyo, 5-1-5 Kashiwanoha, Kashiwa-shi, Chiba 277-8581, Japan
 \and Institute for Materials Research, Tohoku University, 2-1-1 Katahira, Aoba-ku, Sendai 980-8577, Japan
}
\date{
Received: date / Revised version: date}
%
\abstract{
A new experiment to measure vacuum magnetic birefringence (VMB), the OVAL experiment, is reported.
We developed an original pulsed magnet that has a high repetition rate and applies the strongest magnetic field among VMB experiments.
The vibration isolation design and feedback system enable the direct combination of the magnet with a {\fp}cavity.
To ensure the searching potential, a calibration measurement with dilute nitrogen gas and a prototype search for vacuum magnetic birefringence are performed.
Based on the results, a strategy to observe vacuum magnetic birefringence is reported.
} 
\maketitle

\section{Introduction}
\label{chap_intro}
The history of vacuum magnetic birefringence (VMB) dates back to the early 20th century, when Kochel, Euler and Heisenberg derived an effective quantum electrodynamics (QED) Lagrangian at low energies\cite{kochel, hep-th/9812124, Heisenberg1936}.
The Lagrangian predicts that a vacuum itself behaves as a polarizable and magnetizable medium due to virtual particle pair creations and annihilations, which leads to tiny magnetic birefringence under an external magnetic field\cite{PhysRevD.2.2341}.
This is a so-called Cotton-Mouton effect of vacuum and the magnitude can be written as 
\begin{eqnarray}
\mathrm{\Delta} n &=& n_{\parallel} - n_{\perp} \\
&\equiv&k_{\mathrm{CM, vac}}B^2,
\end{eqnarray}
where $n_{\parallel}$ or $n_{\perp}$ is the refractive index parallel or perpendicular to the external magnetic field and $k_{\mathrm{CM, vac}}$ is a coefficient predicted from QED\cite{PhysRevD.2.2341, ADLER1971599}.
At the lowest order in the fine structure constant $\alpha$, the value is given by
\begin{eqnarray}
k_{\mathrm{CM, vac}} &=& \frac{2\alpha^2\hbar^3}{15\mu_0m_e^4c^5}\\
 &=& 4.0 \times 10^{-24}~\mathrm{T}^{-2}.
\end{eqnarray}
In addition to the QED effect, particles predicted by beyond the Standard Model such as axion-like particles\cite{MAIANI1986359, PhysRev.81.899, PhysRevD.37.1237} or millicharged particles\cite{PhysRevLett.97.140402, PhysRevD.75.035011, PhysRevD.12.1132} also induce VMB.
Experimental measurement of VMB is not just a new verification of QED but also a sensitive probe for new particles.

Experimental searches for VMB make use of polarization change.
A linearly polarized light acquires ellipticity $\psi$ after traveling though a birefringent medium, which can be written as
\begin{equation}
\psi = \frac{\pi\mathrm{\Delta} nL}{\lambda}\sin{\theta_{\mathrm{B}}},
\label{eq_ellip}
\end{equation}
where $L$ is the length of the birefringent medium, $\lambda$ is the wavelength of the light, and $\theta_B$ is the angle between the fast axis of the birefringent medium and the polarization axis.

A recent observation of RX J1856.5-3754 suggests a possibility of the first verification of VMB effect\cite{doi:10.1093/mnras/stw2798}.
However, the observed value has large uncertainties on neutron star models and the direction of the neutron magnetization axis.
In order to measure the parameter $k_{\mathrm{CM, vac}}$ and check the result, a well-controlled precise experiment is strongly required.

A tabletop precise scheme to measure VMB is to combine {\fp}cavity with strong a magnetic field, either with a pulsed magnetic field\cite{Battesti2008, refId0, PhysRevA.85.013837} or a static magnetic field\cite{PhysRevD.77.032006, PhysRevD.90.092003, DellaValle2016}.
This kind of experiment can be performed in an experimental room, with all the parameters such as the wavelength of the light, the intensity of the magnetic field, or the length of the magnet being controlled accurately.
An observation of VMB with such kind of experiments provides a chance to precisely investigate QED and theories beyond the Standard Model.
This type of measurement recently gave $k_{\mathrm{CM, vac}} = (-2.4\pm4.8)\times 10^{-23}$ [T$^{-2}$] at 1 $\sigma$ confidence level, a factor of 20 larger sensitivity than that predicted by QED\cite{DellaValle2016}.


In this paper, we report a new apparatus for searching VMB with a pulsed magnet: the OVAL (\textit{Observing VAcuum with Laser}) experiment.
In addition to our stable high finesse {\fp}cavity, a strong and high repeating pulsed magnet features our VMB search.
The magnet has the fastest repetition rate of 0.2 Hz and can apply magnetic field up to 9.0 T.
We successfully combined the strongest magnetic field with a high finesse {\fp}cavity and operated them to measure Cotton-Mouton effect.
The high repetition rate enables faster data acquisition, which results in larger statistics.
A detailed explanation of our prototype apparatus, measurements of dilute low temperature nitrogen and vacuum birefringence, and discussions toward the observation of VMB are described in the following sections.
\section{Experimental Method}
\label{chap_expmeth}
\subsection{Apparatus}
\label{chap_appra}
\begin{figure}[]
\begin{center}
\resizebox{0.37\textwidth}{!}{%
  \includegraphics{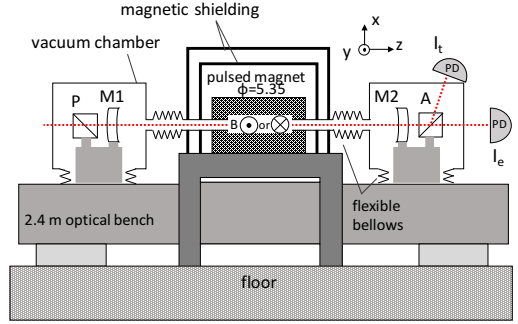}
}
\caption{A schematic of the prototype apparatus.
{\fp}mirrors (M1 and M2), Polarizer (P) and Analyzer (A) are in the vacuum chamber, which is connected to a pulsed magnet.
All the optics are mounted on a optical bench and the magnet is supported from another bench.
See text for more details.
}
\label{fig_drawing}       
\end{center}
\end{figure}

Figure \ref{fig_drawing} represents our prototype setup.
The system is based on a crossed Nicols configuration.
A 1064 nm light from non-planar Nd:YAG ring laser is injected to two perpendicularly placed Glan-Laser prism, Polarizer (P) and Analyzer (A).
Two Photo-Detectors (PDs) are placed after Analyzer to measure the intensity parallel or perpendicular to the incident light, $I_e$ or $I_t$.
A pulsed magnet is placed between the two prisms to induce magnetic birefringence.  
The magnet has an inner diameter $\phi$ of 5.35 mm, which is directly connected to the vacuum chamber, see Sect. \ref{chap_combine} for details.
By rotating the prisms, the angle between the magnetic field and the linear polarization axis $\theta_B$ is aligned to be (45 $\pm$ 1) degrees.
Two mirrors (M1 and M2) are installed in front of and behind the magnet.
The pair of mirrors is called a {\fp}cavity and described in detail in Sect. \ref{chap_fpc}.
All the components including polarizers and {\fp}mirrors are enclosed in the vacuum chamber, which can be evacuated or filled with gas.

Ellipticity signals can be extracted from the intensity changes on the photodiodes.
By denoting the residual static birefringence as $\Gamma$ (see Sect. \ref{chap_fpc} for details) and the ellipticity induced by magnetic field as $\Psi(t)$, the intensity on the detector can be written as
\begin{eqnarray}
I_e &=& I_0(\sigma^2 + (\Gamma + \Psi(t))^2)\\
I_t &=& I_0(1 - \sigma^2 - (\Gamma + \Psi(t))^2),
\end{eqnarray}
where $I_0$ is the transmitted intensity of the {\fp}cavity and $\sigma^2$ is the polarization extinction ratio of the prisms.
Note that the ellipticity is treated as a time dependent effect here.
In our case, $\sigma^2$ is $3\times10^{-7}$ and both $\Gamma^2$ and $\Psi^2$ are smaller than $10^{-4}$.
Therefore, the equation above can be rewritten as
\begin{eqnarray}
\frac{I_e}{I_t} = \sigma^2 + (\Gamma + \Psi(t))^2.
\label{eq_ratio}
\end{eqnarray}

In an actual experiment, one modulates the ellipticity signal $\Psi(t)$ with a known time dependence and searches the same dependence in the measured intensity $I_e(t)$ and $I_t(t)$.
This modulation is achieved by using a pulsed magnet waveform, as will be discussed in Sect. \ref{chap_pulmag}.

\subsection{{\fp}cavity}
\label{chap_fpc}
{\fp}cavity accumulates the light between the two mirrors and thus enhances the effective interaction length with a magnetic field.
The ellipticity acquired in a single path $\psi$ is enhanced as\cite{1367-2630-15-5-053026}
\begin{equation}
\Psi = \frac{2F}{\pi}\psi,
\end{equation}
where $F$ is the finesse of the cavity defined as
\begin{equation}
F=\frac{\pi\sqrt{R}}{1-R-P_{\mathrm{loss}}}.
\label{eq_finesse}
\end{equation} 
Here, $R$ is the reflectivity of the mirrors M1 and M2, and $P_{\mathrm{loss}}$ is the loss inside the cavity other than the mirrors.

Our mirrors are those produced by Advanced Thin Films Inc.\cite{atf}.
The reflectivity is more than 99.999\%, which results in a finesse more than 300~000.
The distance between the two mirrors $L_{\mathrm{cav}}$ is set to be 1.38(1) m.
A feedback control with Pound-Drever-Hall method\cite{doi:10.1063/1.1770414, Drever1983} is applied to lock the laser frequency to the cavity resonance using the PZT modulator on the laser.
In order to achieve extremely stable locking, a forth order low-pass filter is put into the feedback loop.
The resulting unity gain frequency is 70 kHz, with a 60 dB turbulence suppression achieved at a frequency of 1 kHz.
Furthermore, we adopted auto-resonance relocking system.
$I_t$ is monitored to judge whether it is at resonance or not.
When the laser frequency is not resonant with the cavity, feedback control is turned off and the laser frequency is swept by a ramp wave to find the resonance frequency.
When the transmitted intensity $I_t$ jumps up while sweeping, PDH feedback is turned on and feedback control starts.
This system can relock the cavity in less than 1 second.
Though the cavity resonance is rarely broken by the pulsed magnet, this system ensures that the cavity is certainly on resonance when a pulse is shot.

\begin{figure}[]
\resizebox{0.5\textwidth}{!}{%
  \includegraphics{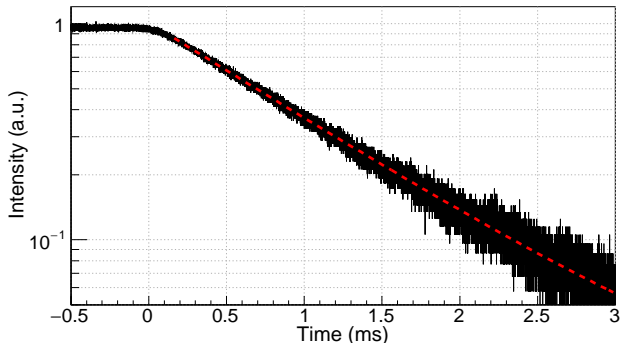}
}
\caption{A typical measurement of the photon lifetime. The feedback control is turned off at 0 ms. Cavity transmitted intensity monitored and fitted with an exponential curve, whose time constant gives photon lifetime. The photon lifetime is 1.0 ms, which gives a finesse of 650~000.}
\label{fig_photonlifetime}       
\end{figure}

The finesse is measured for each pulse cycle by observing the cavity photon lifetime $\tau = FL_{\mathrm{cav}}/\pi c$.
This value can be obtained by measuring the decay of $I_t$ after the feedback is turned off.
Figure \ref{fig_photonlifetime} shows a typical photon lifetime measurement. 
From Fig. \ref{fig_photonlifetime}, combined with the cavity length $L_{\mathrm{cav}}$, the finesse of the cavity is calculated to be 650~000.
Throughout the experiment, the finesse was measured regularly and ranges from 300~000 to 670~000, depending on how long the mirrors had been exposed to the atmosphere.

This long photon lifetime of the cavity acts as a low pass filter and modifies the time dependence of various signals.
First, the dependence of ellipticity on magnetic field square $B^2(t)$ should be modified as\cite{Berceau2010} 
\begin{equation}
\Psi_{\mathrm{LPF}}(t) \propto B^2_{\mathrm{LPF}}(t) = \int^{t}_{-\infty} B^2(t^{\prime})e^{-{t}/{2\tau}} dt^{\prime}.
\end{equation}
In the analysis, this effect is taken into account by recording each waveform of the magnetic field and applying a low pass filter to it.

Another low pass effect is related to the residual birefringence in the cavity $\Gamma$.
Each time the light is reflected on the mirror surface, it acquires ellipticity since the surface is not perfectly isotropic\cite{Bielsa2009}.
Though the polarization change per reflection is quite small and usually negligible, the ultra-high finesse cavity accumulates the effect and thus significantly affects the small ellipticity measurement.
The term $\Gamma$ in Eq. (\ref{eq_ratio}) is dominated by this effect and can be controlled by rotating the two mirrors along the light axis\cite{Berceau2010, Zavattini2006}.
M1 and M2 are both mounted with a rollable mirror mount to control this effect.
Typically the value $|\Gamma|$ ranges from $10^{-6}$ to $10^{-3}$. 

This static birefringence also modifies the time dependence of $I_e/I_t$.
When the birefringence on mirror surface is dominant compared to the polarization extinction ratio ($\Gamma^2 \gg \sigma^2$), the time dependence of $I_e$ becomes\cite{Berceau2010} 

\begin{eqnarray}
I_{e}(t) &=& (\Gamma + \Psi_{\mathrm{LPF}}(t))^2\int^{t}_{-\infty} I_t(t^{\prime})e^{-{t}/{2\tau}} dt^{\prime}\\
&\equiv&(\Gamma + \Psi_{\mathrm{LPF}}(t))^2I_{t,\mathrm{LPF}}(t).
\end{eqnarray}
Thus, in the analysis, we have to first apply low pass filter to $I_t(t)$ and then take the division $I_e(t)/I_{t, \mathrm{LPF}}(t)$ to extract the ellipticity signal.
\subsection{Pulsed magnet}
\label{chap_pulmag}
The importance of magnet in a VMB experiment cannot be overemphasized.
The square of the magnetic field and the length $B^2L$ determine the magnitude of birefringence signal.
The time dependence of the magnetic field determines the detected signal shape.
The operation duty determines the data acquisition rate, i.e.,  statistics.
Advantages of our custom-made pulsed magnet in terms of these three points are described in this section.
See also \cite{Yamazaki2016122, PhysRevLett.118.071803} for details.
\subsubsection{Magnetic field}
\label{chap_magfie}
\begin{figure}[]
\resizebox{0.5\textwidth}{!}{%
  \includegraphics{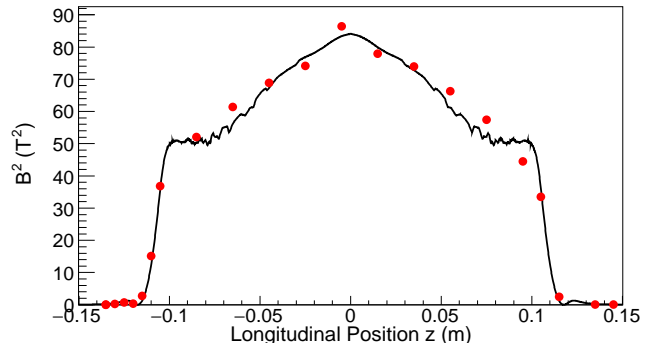}
}
\caption{A profile of the squared magnetic field $B^2(z)$ along the light axis $z$. The solid line is the profile simulated with ANSYS\cite{ansys}. The dots are measured results with a pickup coil.}
\label{fig_magprofile}       
\end{figure}

Our pulsed magnet design is a so-called single racetrack coil.
The magnet is cooled down by liquid nitrogen and a magnetic field of 11.4 T has been achieved at maximum.
In our experiment, the square of magnetic field integrated along the light axis $\int B^2(z) dz$ replaces the term $B^2L$ in Eq. \ref{eq_ellip}.
Figure \ref{fig_magprofile} shows the profile of $B^2(z)$ along the light axis.
This profile is calculated with a finite element simulation software (ANSYS\cite{ansys}) and measured with a pickup coil.
The discrepancy between them is taken into account as an error budget.
By integrating the field, we get,
\begin{equation}
\int B^2(z) dz = 13.8 ~\mathrm{T^2m}.
\end{equation}
This is the largest value used for the VMB experiment ever.
The effective length of magnetic field $L$ is
\begin{equation}
L = \int B^2(z) dz/B^2_{\mathrm{max}} = 0.17~\mathrm{m}.
\end{equation}
\subsubsection{Pulse waveform}
\label{chap_waveform}
As mentioned in Eq. (\ref{eq_ratio}), the magnet also has a role to apply a time dependent modulation to the signal.
Applying a fast modulating magnetic field means a high frequency lock-in detection, thus let the signal avoid noisy background region at lower frequency. 
One strategy taken so far is to rotate permanent or superconductive magnet along the light axis\cite{PhysRevD.77.032006, DellaValle2016}.
The modulation frequency is twice the rotation frequency, which is about 10 Hz.
In our case, the pulsed magnet waveform itself acts as a modulation source, which is as fast as 1 kHz.

\begin{figure}[]
\resizebox{0.5\textwidth}{!}{%
  \includegraphics{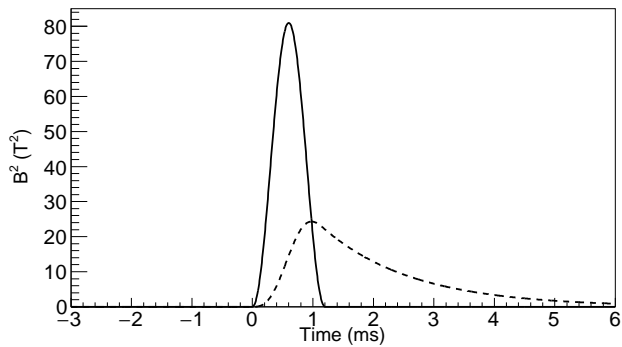}
}
\caption{A waveform of a squared magnetic field $B^2(t)$ (solid line) and that with a low pass filter effect of photon lifetime (dashed line).}
\label{fig_maglpf}       
\end{figure}

The pulse waveform can be written as
\begin{equation}
B^2(t) = B^2_{\mathrm{max}}\sin^2\left({\frac{t}{\sqrt{L_{\mathrm{coil}}C}}}\right),
\end{equation}
where $L_{\mathrm{coil}}$ is the inductance of the magnet coil, which is 49 $\mu$H in our case, and $C$ is the capacitor of the operation bank unit.
The pulse width $\tau_{\textrm{pulse}} = \pi\sqrt{L_{\mathrm{coil}}C}$ can be easily adjusted by changing the capacitance.
$C$ was set to 3.0 mF so that the pulse width becomes 1.2 ms and thus comparable with the cavity photon lifetime.
Figure \ref{fig_maglpf} shows the raw $B^2(t)$ waveform and low pass filtered waveform $B^2_{\mathrm{LPF}}(t)$.

\subsubsection{Fast repeating operation}
\label{chap_repeat}
Our simple magnet design and dedicated operation unit enable to apply alternate magnetic fields with a high repetition rate.
We first apply a pulsed magnetic field with 9.0 T, then apply the second one with 4.5 T in two seconds, with the direction reversed. 
We denote the first one as $+$9.0 T and second one as $-$4.5 T explicitly.
We call the successive $+$9.0 T and $-$4.5 T shots in a set as a ``cycle.''
Since the birefringence signal is proportional to $B^2$, a noise cancellation can be achieved by adding one to the other.

The biggest advantage of our magnet is the high repetition rate $f_{\mathrm{rep}}$.
In previous VMB search experiments, the sensitivities are limited by noise proportional to statistics.
A higher repetition rate results in fast data acquisition and better sensitivity.
Usually, the repetition rate of a pulsed magnet is limited both by the time to recharge capacitor bank and the time to cool down the pulsed magnet.
Our new design of reduced thermally resistive materials with thin nonmagnetic stainless steel support structure effectively increases cooling efficiency.
Combined with the operation unit above, repeating operation of two pulses with $+$9.0 T and $-$4.5 T in 10 seconds is achieved, which results in a operation at 0.17 Hz.

\subsection{Combination of a pulsed magnet with {\fp}cavity}
\label{chap_combine}
The pulsed magnet pipe is directly conneccted to the vacuum chamber.
The magnetic field region has to be confined as small as possible in order to generate higher magnetic field effectively.
The direct connection results in compressing magnetic field significantly.  
On the other hand, the cavity requires that the intra-cavity beam should not be cut by the magnet pipe.
Also, since {\fp}cavity is quite easily disturbed by outer turbulence, it is extremely important to isolate vibration from the pulsed magnet. 
To achieve both direct connection and isolation of vibration, we design and assemble the apparatus as follows.
\begin{figure}[]
\resizebox{0.5\textwidth}{!}{%
  \includegraphics{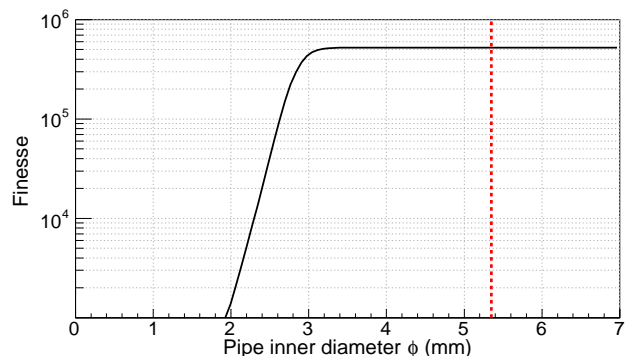}
}
\caption{The dependence of the {\fp}finesse on the pipe inner diameter $\phi$. The inner diameter of our magnet is shown by the dotted line.}
\label{fig_pipeloss}       
\end{figure}

First, the minimum requirement for the pipe diameter is estimated by calculating cavity intra-beam loss due to magnet pipe.
From the curvature of the mirror (2 m) and cavity length ($L_{\mathrm{cav}}=1.38$ m), the beam waist of TEM$_{00}$ mode $w_0$ is calculated as 0.55 mm.
The loss due to the pipe is calculated as,
\begin{equation}
P_{\mathrm{loss}}=1-\int_{0}^{\frac{\phi}{2}}\frac{4r}{w_0^2}\exp\left(-\frac{2r^2}{w_0^2}\right)dr.
\end{equation}
By substituting this loss $P_{\mathrm{loss}}$ in Eq. (\ref{eq_finesse}), we can estimate the dependence of finesse on the pipe inner diameter.
Figure \ref{fig_pipeloss} presents the requirement for pipe inner diameter for a finesse of 500~000.
Thus, the magnet pipe is designed with an inner diameter $\phi$ of 5.35 mm, which gives a margin of 2 mm in diameter.

\begin{figure}[]
\resizebox{0.5\textwidth}{!}{%
  \includegraphics{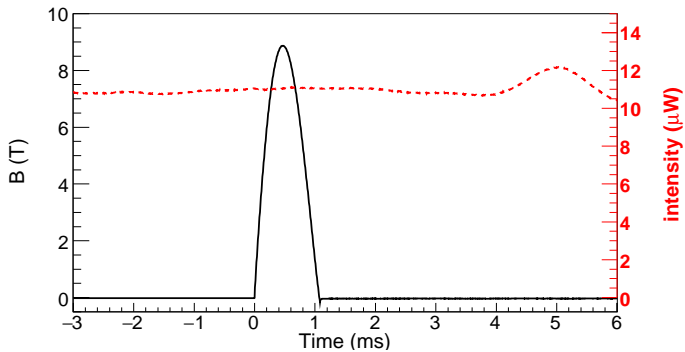}
}
\caption{The intensity of the cavity transmitted light when a $+$9.0 T magnetic field is applied. The disturbance is later than 4 ms after a pulsed field is applied.}
\label{fig_Tur}       
\end{figure}

Second our apparatus is assembled as shown in Fig. \ref{fig_drawing}.
Flexible bellows are inserted in two connection points, one between the pulsed magnet and the vacuum chamber and the other between the vacuum chamber and the optical bench.
M1, M2, P and A are mounted from the optical bench and detached from the inner wall of the vacuum chamber.
In addition to this design, multilayer magnetic shieldings reduce electromagnetic disturbance from the pulsed magnet.
The stable feedback system with the strong turbulence supression at a low frequency also contributes the robustness of the cavity.
Figure \ref{fig_Tur} shows the signal of $I_t$ when a $+$9.0 T pulsed magnetic field is generated.
No apparant effect on the transmitted intensity can be seen while magnetic field is applied.

Table \ref{tab_1} summarizes our experimental parameters for the prototype measuremnet, see the Sect. \ref{chap_measurement} for details.
The unique apparatus gives the highest magnetic field with {\fp}cavity ever.

\begin{table}
\caption{Experimental parameters used in the prototype experiment compared with other VMB experiments\cite{DellaValle2016, refId0}}
\label{tab_1}       
\begin{center}
\begin{tabular}{l|ccc}
\hline\noalign{\smallskip}
Parameters & PVLAS & BMV & OVAL  \\
\noalign{\smallskip}\hline\noalign{\smallskip}
Finesse &700~000 & 450~000 & 320~000 \\
Magnet type&static & pulsed & pulsed \\
Maximum $B$ (T)&2.5 & 6.5 & 9.0 \\
Length $L$ (m)&1.6 & 0.137 & 0.17 \\
$f_{\mathrm{rep}}$ (Hz)&--- & 0.0017 & 0.17 \\
$\Psi$ (rad)&$2.6 \times 10^{-11}$ & $2.0 \times 10^{-11}$ & $2.4 \times 10^{-11}$ \\
\noalign{\smallskip}\hline
\end{tabular}
\end{center}
\end{table}
\section{Measurement}
\label{chap_measurement}
\subsection{Calibration}
\label{chap_calib}
To calibrate the sensitivity, a measurement with dilute molecule nitrogen gas is performed.
The vacuum chamber is filled with dry nitrogen gas with the pressure measured at the both sides of the chamber, ranging from 100 to 400 Pa.

Since our magnet is directly connected to the vacuum chamber and cooled down to 77 K, the inner gas temperature is also far below the room temperature.
Considering that the pressure range is in viscous flow, the inner gas temperature can be assumed to be equal to the stainless tube of the magnet, as low as 77 K.
The upper value of the temperature is estimated from the temperature of magnet coil during repeating operation, which is heated by Joule heat up to 95 K.
Thus the gas temperature is estimated to be (86 $\pm$ 9) K.
This calibration measurement is not only a verification of our apparatus, but also the first measurement of nitrogen gas birefringence at low temperature.

One another factor to be taken into account is the Faraday rotation of nitrogen gas\cite{QUA:QUA560010518}. 
The effect arises from the presence of longitudinal magnetic field.
Faraday rotation is the rotation of the major axis of the ellipse, whose size $\Theta_{\mathrm{F}}$ is given by
\begin{equation}
\Theta_{\mathrm{F}}=\frac{2Fk_{\mathrm{F}}B_{\parallel}L}{\lambda},
\end{equation}
where $k_{\mathrm{F}}$ is the Faraday rotation parameter of nitrogen.
Just as birefringence, the integration of longitudinal magnetic field $\int B_{\parallel}(z) dz$ contributes $\Theta_{\mathrm{F}}$ and the value is estimated to be $0.24(1)~\mathrm{Tm}$ with the same procedure described in Sect. \ref{chap_pulmag}.
In addition, static Faraday rotation on the {\fp}mirror surface $\epsilon$ is also considered.
The effect can be controlled with mirror rolling, typically 10 times smaller than $|\Gamma|$.
The total intensity behavior in nitrogen measurement is given by
\begin{equation}
\frac{I_e(t)}{I_{t, \mathrm{LPF}}(t)} = (\Gamma + \Psi_{\mathrm{LPF}}(t))^2 + (\epsilon + \Theta_{\mathrm{F, LPF}}(t))^2
\label{eq_withfaraday}
\end{equation}

Since Faraday effect depends on magnetic field linearly while Cotton-Mouton effect square, these two effects can be separated from each other by combining the $+9.0$ T and $-4.5$ T data.

\begin{figure}[]
\resizebox{0.5\textwidth}{!}{%
  \includegraphics{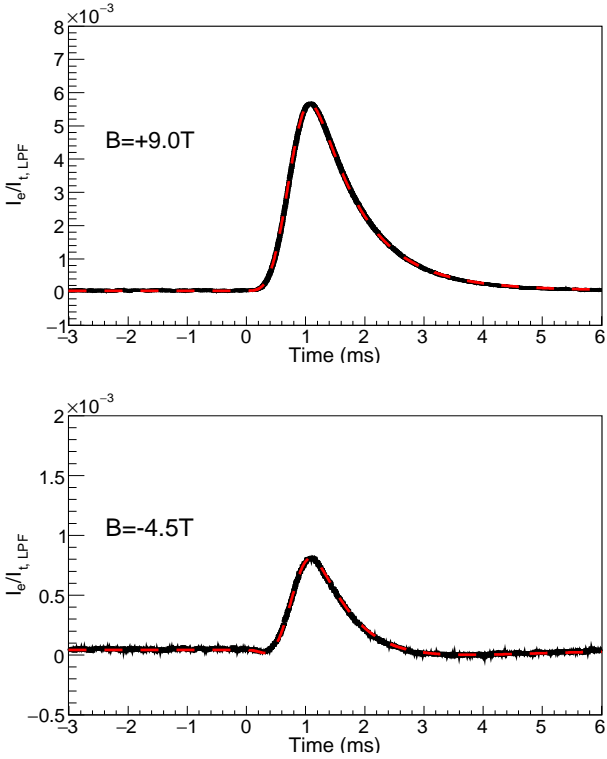}
}
\caption{A typical polarization change signal with a magnetic field of $+$9.0 T and $-$4.5 T. The best fit result with Eq. (\ref{eq_withfaraday}) are also overdrawn (dashed line).}
\label{fig_N2fit}       
\end{figure}

The pressure is set at 4 values and 10 cycles of data were acquired at each pressure.
Figure \ref{fig_N2fit} shows signals with $+$9.0 T and $-$4.5 T.
The two signals are fitted simultaneously with the four parameters, $\Gamma$, $k_{\mathrm{CM}}(\mathrm{N_2})$, $\epsilon$, and $k_{\mathrm{F}}(\mathrm{N_2})$ being free.
Residual fluctuation can be canceled almost perfectly by taking the division $I_e/I_{t, \mathrm{LPF}}$.

\begin{figure}[]
\resizebox{0.5\textwidth}{!}{%
  \includegraphics{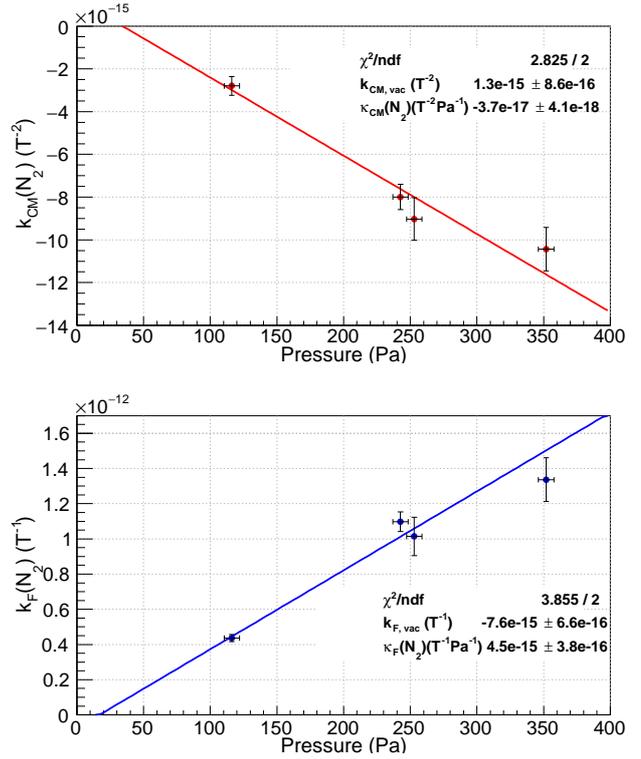}
}
\caption{A plot of measured $k_{\mathrm{CM}}(\mathrm{N_2})$ and $k_{\mathrm{F}}(\mathrm{N_2})$ at each pressure. The results are fitted with Eq. (\ref{fitting1}, \ref{fitting2}). The offsets $k_{\mathrm{CM, vac}}$ and $k_{\mathrm{F, vac}}$ are consistent with zero.}
\label{fig_pressure}       
\end{figure}

This fitting was performed for all the data taken at each pressure. 
The average of the fitting center values and its standard error give the magnitude of Cotton-Mouton effect and Faraday effect at each pressure.
The pressure dependence of these effects are shown in Fig. \ref{fig_pressure}. 
Overwritten lines are fittings by
\begin{eqnarray}
\label{fitting1}
k_{\mathrm{CM}}(\mathrm{N_2}) &=& \kappa_{\mathrm{CM}}(\mathrm{N_2})\times P + k_{\mathrm{CM, vac}}\\
k_{\mathrm{F}}(\mathrm{N_2}) &=& \kappa_{\mathrm{F}}(\mathrm{N_2})\times P + k_{\mathrm{F, vac}},
\label{fitting2}
\end{eqnarray}
 where $k_{\mathrm{F, vac}}$ is Faraday rotation parameter of vacuum and should be zero.
From this measurement, the parameters $\kappa_{\mathrm{CM}}$ and $\kappa_{\mathrm{F}}$ can be obtained.
Note that the sign of $\Gamma$ and $\epsilon$ can be determined by rotating polarizers.
The fitting results are
\begin{eqnarray}
\kappa_{\mathrm{CM}}(\mathrm{N_2}) &=& (-3.1 \pm 0.4 \pm 0.1 )\times10^{-17} (\mathrm{T^{-2}Pa^{-1}})\\
\kappa_{\mathrm{F}}(\mathrm{N_2}) &=& (4.5 \pm 0.4 \pm 0.1)\times10^{-15} ~(\mathrm{T^{-1}Pa^{-1}}),
\end{eqnarray}
where the first uncertainty is the fitting uncertainty and the second is the systematic uncertainties summarized in Table \ref{tab_2}. 
\begin{table}[]
\caption{Error budgets on the nitrogen gas measurement.}
\label{tab_2}       
\begin{center}
\begin{tabular}{c|cc}
\hline\noalign{\smallskip}
 &\multicolumn{2}{c}{Relative uncertainty}\\
Parameter &$\kappa_{\mathrm{CM}}(\mathrm{N_2})$ & $\kappa_{\mathrm{F}}(\mathrm{N_2})$ \\
\noalign{\smallskip}\hline\noalign{\smallskip}
Finesse &\multicolumn{2}{c}{$4\times 10^{-2}$} \\
$\int B^2(z) dz$&$6\times 10^{-2}$ & --- \\
$\int B_{\parallel}(z) dz$&--- & $8\times 10^{-2}$ \\
$\theta_{B}$& $1.5\times 10^{-2}$ &--- \\
Guiding efficiencies of $I_e$ and $I_t$& \multicolumn{2}{c}{$4\times 10^{-2}$}\\
Laser wavelength $\lambda$& \multicolumn{2}{c}{$<1\times 10^{-4}$} \\
Pressure &\multicolumn{2}{c}{considered in Fig. \ref{fig_pressure}} \\
\noalign{\smallskip}\hline
\end{tabular}
\end{center}
\end{table}
The finesse is measured between each pulse and the drift is taken as the error.
The error for magnetic field is estimated as described in Sect. \ref{chap_magfie}.
$\theta_{\mathrm{B}}$ is aligned to (45 $\pm$ 1) degrees with a jig.
Guiding efficiencies are estimated with a power meter.
The laser wavelength is determined from the oscillation wavelength of Nd:YAG laser.
Inner gas pressure is measured at both chambers and considered in the fitting, see Fig. \ref{fig_pressure}.

\begin{figure}[]
\resizebox{0.5\textwidth}{!}{%
  \includegraphics{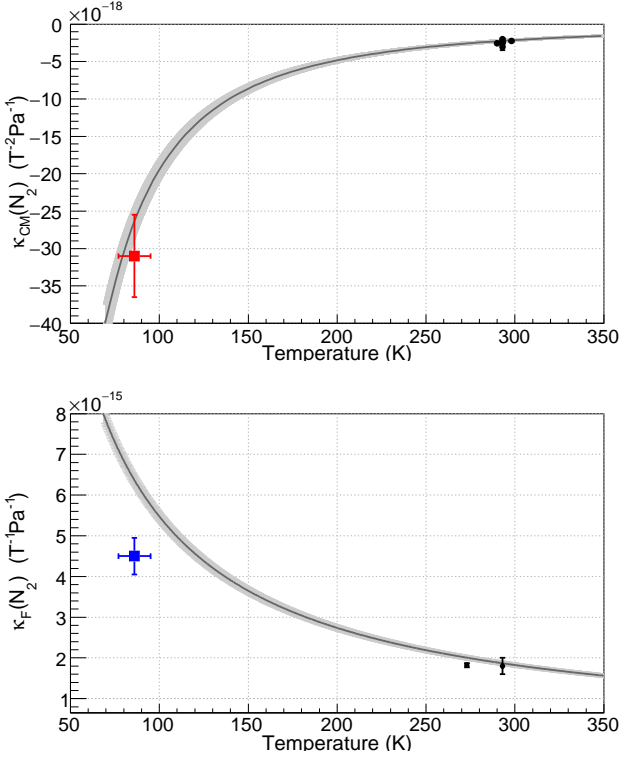}
}
\caption{Comparison of the measured results (rectangle) with theoretical line. The gray region corresponds to theoretical uncertainty. Previous results are also drawn with dots. See\cite{doi:10.1080/014423597230316} and references there for birefringence experiments and\cite{Ingersoll:54, Ingersoll:56, Ingersoll:58, QUA:QUA560010518} for Faraday effect measurements.}
\label{fig_temp}       
\end{figure}

In order to compare the results with the theoretical value, one must take into account the dependence on the gas temperature.
It derives from the density of ideal gas that is proportional to the inverse of the temperature and the axial molecular rotation structure.
The latter one for birefringence parameter $\kappa_{\mathrm{CM}}$ are widely discussed, for example in \cite{doi:10.1080/014423597230316}, while that for Faraday rotation is usually neglected.
Since the major purpose here is to calibrate and verify the operation of our apparatus, we adopted the theoretical prediction from \cite{doi:10.1080/014423597230316} for birefringence and simple inverted dependence on temperature for Faraday rotation\cite{Ingersoll:54, Ingersoll:56, Ingersoll:58}. 
Figure \ref{fig_temp} compares the result with theoretical calculation in terms of temperature.

The good match for birefringence result assures the ability to detect small polarization change in our apparatus.

\subsection{Vacuum magnetic birefringence measurement}
\label{chap_vmbmea}
After calibrating and verifying the apparatus, we performed a vacuum magnetic birefringence search with the intensity $I_0$ equal to 10 $\mu$W, a finesse $F$ equal to 320~000 and $\Gamma$ set to be 2.2 $\times$ $10^{-3}$.
The pressure inside pulsed magnet was estimated to be lower than $10^{-3}$ Pa, which is low enough in current experimental sensitivity.
The data acquisition scheme is totally same with that in nitrogen measurement.
100 cycles data were collected, each containing both $+$9.0 T and $-$4.5 T data.
Note that due to our fast repetition operation, the time for data taking $T_{\mathrm{run}}$ is as short as 15 minutes.
Drifts such as temperature changes, inner pressure changes, or long term misalignment can be neglected in this measurement.

This time, since the pressure is considerably low, the square of ellipticity $\Psi^2$ and Faraday rotation $\Theta_{\mathrm{F}}^2$ can be neglected.
The ellipticity signal can be extracted by 
\begin{equation}
\left.\frac{I_e}{I_t}\right|_{+} + \frac{9.0}{4.5}\left.\frac{I_e}{I_t}\right|_{-}= \left(1+\frac{9.0}{4.5}\right)\times\left(\sigma^2 + \Gamma^2 + 2\Gamma\Psi(t) + \epsilon^2\right),
\label{eq_vacellip}
\end{equation}
where the subscripts $+$ and $-$ express $+9.0$ T and $-4.5$ T. 
Note that $\sigma^2$ and $\Gamma^2$ can be measured when no magnetic field is applied and static ellipticity $\epsilon^2$ is much smaller than $\Gamma^2$. 
We fit the signal with a time independent term and $B^2_{\mathrm{LPF}}(t)$ term.

\begin{figure}[]
\resizebox{0.5\textwidth}{!}{%
  \includegraphics{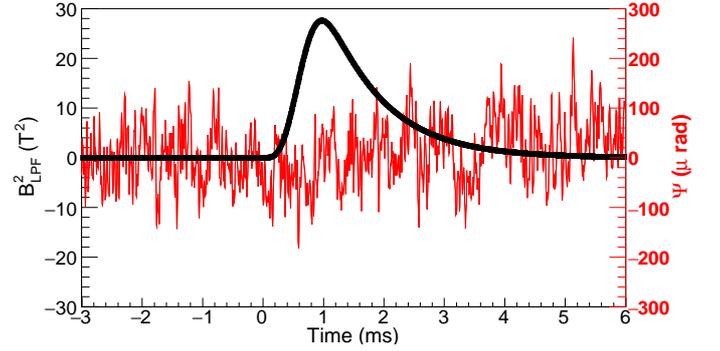}
}
\caption{An typical ellipticity measurement in vacuum. The black line presents magnetic field squared and the red line presents measured ellipticity.}
\label{fig_Vac}       
\end{figure}

\begin{figure}[]
\resizebox{0.5\textwidth}{!}{%
  \includegraphics{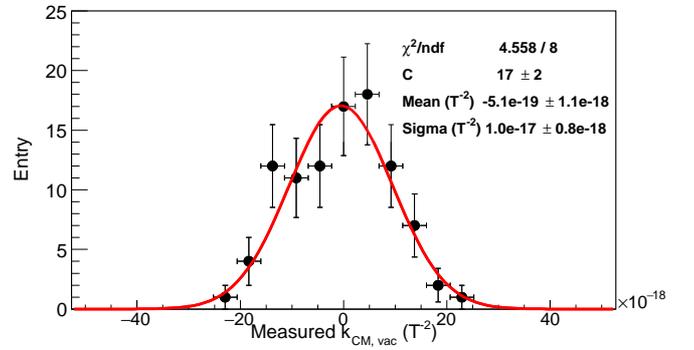}
}
\caption{A histogram of the $k_{\mathrm{CM, vac}}$ fitting center value for 100 cycles. The mean and its error of fitted Gaussian give a preliminary limit on $k_{\mathrm{CM, vac}}$.}
\label{fig_vachist}       
\end{figure}

Figure \ref{fig_Vac} shows a waveform of typical ellipticity measurement.
By fitting the signal and extracting the best fitting result for each cycle, we can make a distribution of the center value in a histogram, as shown in Fig. \ref{fig_vachist}.
Then the distribution was fitted with Gaussian, whose mean value and error give a limit for vacuum magnetic birefringence parameter, as
\begin{equation}
k_{\mathrm{CM, vac}} = (-0.5 \pm 1.1) \times 10^{-18} ~(\mathrm{T}^{-2}).
\end{equation}

Since the run time $T_{\mathrm{run}}$ is only 15 minutes, systematic error such as a drift of finesse or the stability of magnetic field is much smaller than statistical error.

\section{Prospects}
\label{chap_prosp}
The sensitivity of our prototype measurement is limited by shot noise.
The dependence of sensitivity on various factors is as follows.
\begin{equation}
\Delta k_{\mathrm{CM, vac}} \propto \frac{1}{B^2LF\sqrt{I_0T_{\mathrm{DAQ}}}},
\end{equation}
where $T_{\mathrm{DAQ}}$ is the effective data acquisition time proportional to $T_{\mathrm{run}}$, $f_{\mathrm{rep}}$, and $\tau_{\textrm{pulse}}$.
The parameters in the prototype measurement are $F=320~000$, $B^2L=13.8~\mathrm{T^2m}$, and $I_0=10~\mu$W.
To reach the sensitivity to observe QED predicted VMB, we are upgrading the apparatus.

The largest improvement can be achieved by increasing the cavity transmitted power $I_0$.
In the prototype measurement, this was limited simply by the input power of the laser.
The laser has been already upgraded and 10 mW transmitted intensity $I_0$ will be achieved.
In addition, new mirrors are prepared and a finesse of 650~000 can be achieved, which is 2 times higher than the prototype measurement.
They contribute a factor of 60 times better sensitivity.

The upgrade of pulsed magnet is also ongoing.
The next generation pulsed magnet is designed to reach 15 T, with a new Ag-Cu wire coil. 
By placing 4 magnets between {\fp}mirrors, the interaction length can also be extended four times longer.
The inductance of coil $L_\mathrm{coil}$ will also be increased four times.
In addition, the capacitance $C$ will be upgrade to 12.0 mF, this contributes to a 4 times longer pulse width $\tau_{\textrm{pulse}}$.
Though the repetition rate will be decresed to 0.1 Hz, these upgrades contribute another 40 times better sensitivity.

Both the upgrade of {\fp}cavity and pulsed magnet will increse the theoretical sensitivity up to
\begin{equation}
\Delta k_{\mathrm{CM, expected}} = \frac{1.2 \times 10^{-20}}{\sqrt{T_{\mathrm{run}}\mathrm{~(sec.)}}}~ (\mathrm{T^{-2}}).
\end{equation}
This will be the best sensitivity ever for VMB search.
Thus, with the sensitivity above, the first observation of VMB with talbetop apparatus will be achieved in 3 months run time.
\section{Conclusion}
\label{chap_conclu}
The status of a new experiment, the OVAL experiment, aiming to observe VMB is described.
New high repetition strong pulsed magnets are developed and combined with a {\fp}cavity.
Our stable {\fp}cavity makes it possible to connect a pulsed magnet to the vacuum chamber directly and apply higher magnetic field.
These features provide the largest birefringence signal.
In addition, the highest repetition rate of the pulsed magnet provides higher statistics.
A calibration measuremnet was performed and shows the ability of the apparatus to measure small birefringence.
Also, a prototype vacuum measurement is performed and shot noise limited sensitivity is achieved.
Based on these results, a future upgrade plan of our apparatus is established.
Upgrade of both the {\fp}cavity and the pulsed magnet are now ongoing.
The future sensitivity will be the best one ever achieved and enable the first observation of VMB.

\section*{Acknowledgement}
\label{chap_thanks}
We thank C. Rizzo, N. Mio, M. Daimon, M. Ando, A. Ishida, A. Sugamoto, K. Yamashita, Y. Enomoto, and Y. Uesugi for fruitful discussions and support. This experiment was supported by Advanced Photon Science Aliance. This research is a part of Inter-university Cooperative Research Program of the Institute for Materials Research, Tohoku University (Proposals No. 14K0018 and No. 15K0080). This work was supported by JPSP KAKENHI Grant Number JP16H03970, JP17H05398 and MEXT KAKENHI Grant Number JP26104701.
\bibliographystyle{epj}
\bibliographystyle{apsrev4-1}

\end{document}